\documentclass{ws-ijmpa}

\begin{document}

\markboth{Colin Morningstar}
{Exploring the spectrum of QCD}

%
\catchline{}{}{}{}{}
%

\title{EXPLORING THE SPECTRUM OF QCD\\ USING A SPACE-TIME LATTICE}

\author{COLIN MORNINGSTAR}

\address{Department of Physics,
         Carnegie Mellon University,
         Pittsburgh, PA 15213, USA}

\maketitle

\begin{history}
\received{September 21, 2005}
\end{history}

\begin{abstract}
Some past and ongoing explorations of the spectrum of QCD using
Monte Carlo simulations on a space-time lattice are described.
Glueball masses in the pure-gauge theory are reviewed, and the energies
of gluonic excitations in the presence of a static quark-antiquark pair
are discussed.  Current efforts to compute the baryon spectrum
using extended three-quark operators are also presented, emphasizing
the need to use irreducible representations of the cubic point group
to identify spin quantum numbers in the continuum limit.
\end{abstract}

\ccode{PACS numbers: 12.38.Gc, 11.15.Ha, 12.39.Mk}

\section{Introduction}

Spectroscopy is a powerful tool for distilling the key degrees of
freedom in a given system.  At the present time, the best way of
extracting the spectrum of states from the QCD Lagrangian is
Monte Carlo computer calculations using a space-time lattice.
A spectrum determination requires the extraction of many excited-state
energies, so a brief discussion on how excited-state energies can be
determined from Monte Carlo estimates of correlation functions
in Euclidean field theory is warranted.  In this talk, several past
explorations of the QCD spectrum are outlined, in particular, 
glueball masses in the pure-gauge theory
and the energies of gluonic excitations of the so-called static
quark-antiquark potential.  Also, ongoing efforts by the Lattice
Hadron Physis Collaboration (LHPC) to determine the baryon spectrum
using extended three-quark operators are described.

\section{Extracting Excited-State Energies and Resonances}

The Monte Carlo method can be applied to obtain estimates of the path
integrals which yield a Hermitian matrix of correlation functions
$C_{ij}(t)=\langle 0\vert\ O_i(t)\overline{O}_j(0)\ \vert 0\rangle$,
where $\overline{O}_j(0)$ creates the states of interest at time
$t=0$ and $O_i(t)$ annihilates such states at a later time $t$.
The procedure for extracting the lowest energies $E_0,E_1,E_2,\dots$ from
this matrix is well known\cite{cmichael,luscherwolff}.
Let $\lambda_n(t,t_0)$ denote the eigenvalues of the hermitian matrix 
$C(t_0)^{-1/2}\,C(t)\,C(t_0)^{-1/2}$, where $t_0$ is some fixed reference time
(typically small) and the eigenvalues, also known as the {\em principal} 
correlation functions, are ordered such that $\lambda_0\geq\lambda_1\geq\cdots$ 
as $t$ becomes large.  Then one can show that
\begin{eqnarray}
 \lim_{t\rightarrow\infty}\lambda_n(t,t_0) &=& e^{-E_n (t-t_0)}\Bigl(
  1 + O(e^{-\Delta_n (t-t_0)})\Bigr),\\
  \Delta_n &=& \min_{k\neq n}\vert E_k-E_n\vert.
\end{eqnarray}
Determinations of the principal correlators $\lambda_n(t,t_0)$ for 
large temporal separations $t$ yield the desired energies $E_n$.
Since statistical fluctuations grow with increasing $t$, it is
crucial that contributions from higher-lying states be diminished so
that the desired lowest-lying energies dominate the principal correlators
well before the signal-to-noise ratio falls.  Judiciously chosen
quark-field and gluon-field smearings is one important ingredient for
reducing couplings to the short wavelength modes of the theory.  The use
of large sets of extended operators is another key ingredient.

Our Monte Carlo calculations are carried out in
a finite-sized box with periodic boundary conditions.  Given the
finite volume and the discrete nature of the allowed momenta in
the box, the masses and widths of resonances (unstable hadrons) cannot 
be calculated directly, but must be deduced from the discrete spectrum
of finite-volume stationary states for a range of box 
sizes\cite{dewitt,wiese88,luscher91B,rum95}. Such applications
require prohibitive computational resources, but our goal in the 
baryon project is to obtain a first exploratory scan of the
spectrum, so simply obtaining the finite-volume spectrum for a
few judiciously-chosen volumes should suffice for ferreting out the hadron
resonances from the less interesting scattering states.

\section{Glueballs and Gluonic Excitations in Presence of Static 
$\overline{Q}Q$ Pair}

\begin{figure}[t]
\centerline{\psfig{file=fig1A.eps,width=6.25cm}
\psfig{file=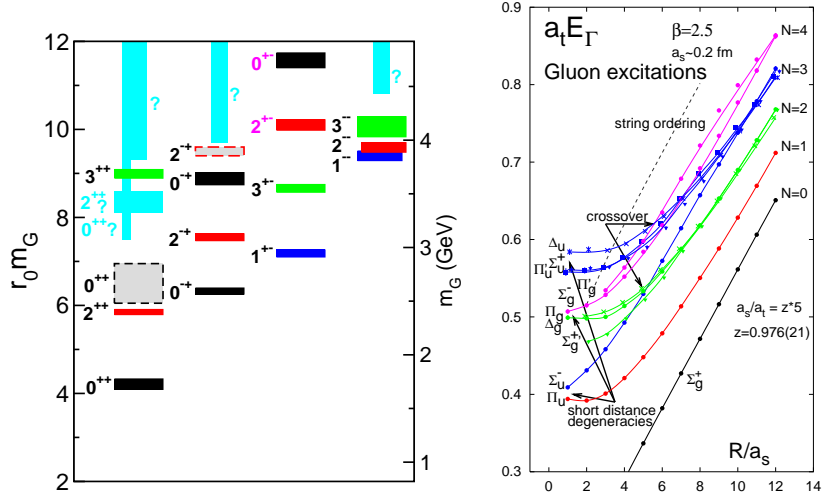,width=4.75cm}}
\vspace*{8pt}
\caption{The glueball mass spectrum\protect\cite{glueballs} in the pure-gauge
theory (left).  The spectrum of gluonic excitations\protect\cite{jkm} in the 
presence of a static quark-antiquark pair separated by distance $R$ 
(right).\label{fig:one}}
\end{figure}

Glueball masses in the pure-gauge theory, shown in Fig.~\ref{fig:one},
were the first exploration of the QCD spectrum\cite{glueballs} in which 
I was involved. The spectrum can be fairly well described by a bag model
description of the gluons, whereas string models seem to fare less well.
Inclusion of light-quark effects is an ongoing challenge.

The issue of string formation of the gauge-field in the presence of a static
quark-antiquark pair was my next exploration\cite{jkm}.  For quark-antiquark
separations $R$ greater than 2 fm, the spectrum of gluonic excitations 
agreed without exception with that expected from an effective string
theory description of the gauge field (see Fig.~\ref{fig:one}), and
a fine structure provided tantalizing clues to the nature of such an
effective string theory.  A dramatic level reordering was observed
as $R$ became smaller, suggesting a bag model picture or multipole expansion
may be more
relevant at such scales.  These energies were used\cite{jkm2} as a
starting point in a Born-Oppenheimer treatment of heavy-quark mesons,
both conventional and hybrid (bound by an excited gluon field).  Agreement
of level splittings from direct Monte Carlo calculations with those
in the leading Born-Oppenheimer approximation validated such a treatment
and provided a compelling physical picture of both conventional and hybrid
heavy-quark mesons.

\section{Baryons}

\begin{figure}[b]
\centerline{
\raisebox{0mm}{\setlength{\unitlength}{1mm}
\thicklines
\begin{picture}(16,10)
\put(8,6.5){\circle{6}}
\put(7,6){\circle*{2}}
\put(9,6){\circle*{2}}
\put(8,8){\circle*{2}}
\put(4,0){single-}
\put(5,-3){site}
\end{picture}} 
\raisebox{0mm}{\setlength{\unitlength}{1mm}
\thicklines
\begin{picture}(16,10)
\put(7,6.2){\circle{5}}
\put(7,5){\circle*{2}}
\put(7,7.3){\circle*{2}}
\put(14,6){\circle*{2}}
\put(9.5,6){\line(1,0){4}}
\put(4,0){singly-}
\put(2,-3){displaced}
\end{picture}}  
\raisebox{0mm}{\setlength{\unitlength}{1mm}
\thicklines
\begin{picture}(20,8)
\put(12,5){\circle{3}}
\put(12,5){\circle*{2}}
\put(6,5){\circle*{2}}
\put(18,5){\circle*{2}}
\put(6,5){\line(1,0){4.2}}
\put(18,5){\line(-1,0){4.2}}
\put(6,0){doubly-}
\put(4,-3){displaced-I}
\end{picture}}  
\raisebox{0mm}{\setlength{\unitlength}{1mm}
\thicklines
\begin{picture}(20,13)
\put(8,5){\circle{3}}
\put(8,5){\circle*{2}}
\put(8,11){\circle*{2}}
\put(14,5){\circle*{2}}
\put(14,5){\line(-1,0){4.2}}
\put(8,11){\line(0,-1){4.2}}
\put(4,0){doubly-}
\put(1,-3){displaced-L}
\end{picture}}   
\raisebox{0mm}{\setlength{\unitlength}{1mm}
\thicklines
\begin{picture}(20,12)
\put(10,10){\circle{2}}
\put(4,10){\circle*{2}}
\put(16,10){\circle*{2}}
\put(10,4){\circle*{2}}
\put(4,10){\line(1,0){5}}
\put(16,10){\line(-1,0){5}}
\put(10,4){\line(0,1){5}}
\put(4,0){triply-}
\put(1,-3){displaced-T}
\end{picture}} 
\raisebox{0mm}{\setlength{\unitlength}{1mm}
\thicklines
\begin{picture}(20,12)
\put(10,10){\circle{2}}
\put(6,6){\circle*{2}}
\put(16,10){\circle*{2}}
\put(10,4){\circle*{2}}
\put(6,6){\line(1,1){3.6}}
\put(16,10){\line(-1,0){5}}
\put(10,4){\line(0,1){5}}
\put(4,0){triply-}
\put(2,-3){displaced-O}
\end{picture}}  }
\vspace*{8pt}
\caption{The spatial arrangments of the extended three-quark baryon
operators used\protect\cite{baryons}. Smeared quark fields are
shown by solid circles, line segments indicate
gauge-covariant displacements, and each hollow circle indicates the location
of a Levi-Civita color coupling.  For simplicity, all displacements
have the same length in an operator.
\label{fig:two}}
\end{figure}
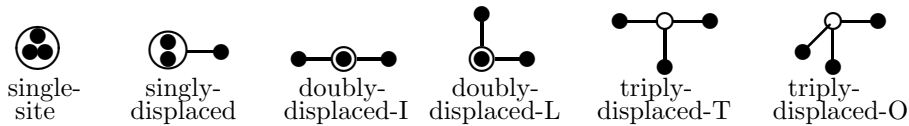

I am currently a member of the Lattice Hadron Physics Collaboration (LHPC).
One of our goals is to compute the spectrum of baryon resonances,
with an eye towards later determining the meson spectrum.  We have
designed large sets of extended, gauge-invariant three-quark operators
to facilitate this task\cite{baryons}.  The usual operator construction 
which mimics the approach one would take in continuous space-time is very
cumbersome, especially when tackling an entire spectrum.  Our approach 
combines the
physical characteristics of baryons with the symmetries of the lattice
regularization of QCD used in simulations.  For baryons at rest, our 
operators are formed using group-theoretical projections onto the irreducible
representations (irreps) of the $O_h$ symmetry group of a three-dimensional cubic lattice.
There are four two-dimensional irreps $G_{1g}, G_{1u}, G_{2g}$, $G_{2u}$
and two four-dimensional representations $H_g$ and $H_u$. 
The continuum-limit spins $J$ of our states can be deduced by examining
degeneracy patterns across the different $O_h$ irreps.

\begin{figure}[t]
\centerline{\psfig{file=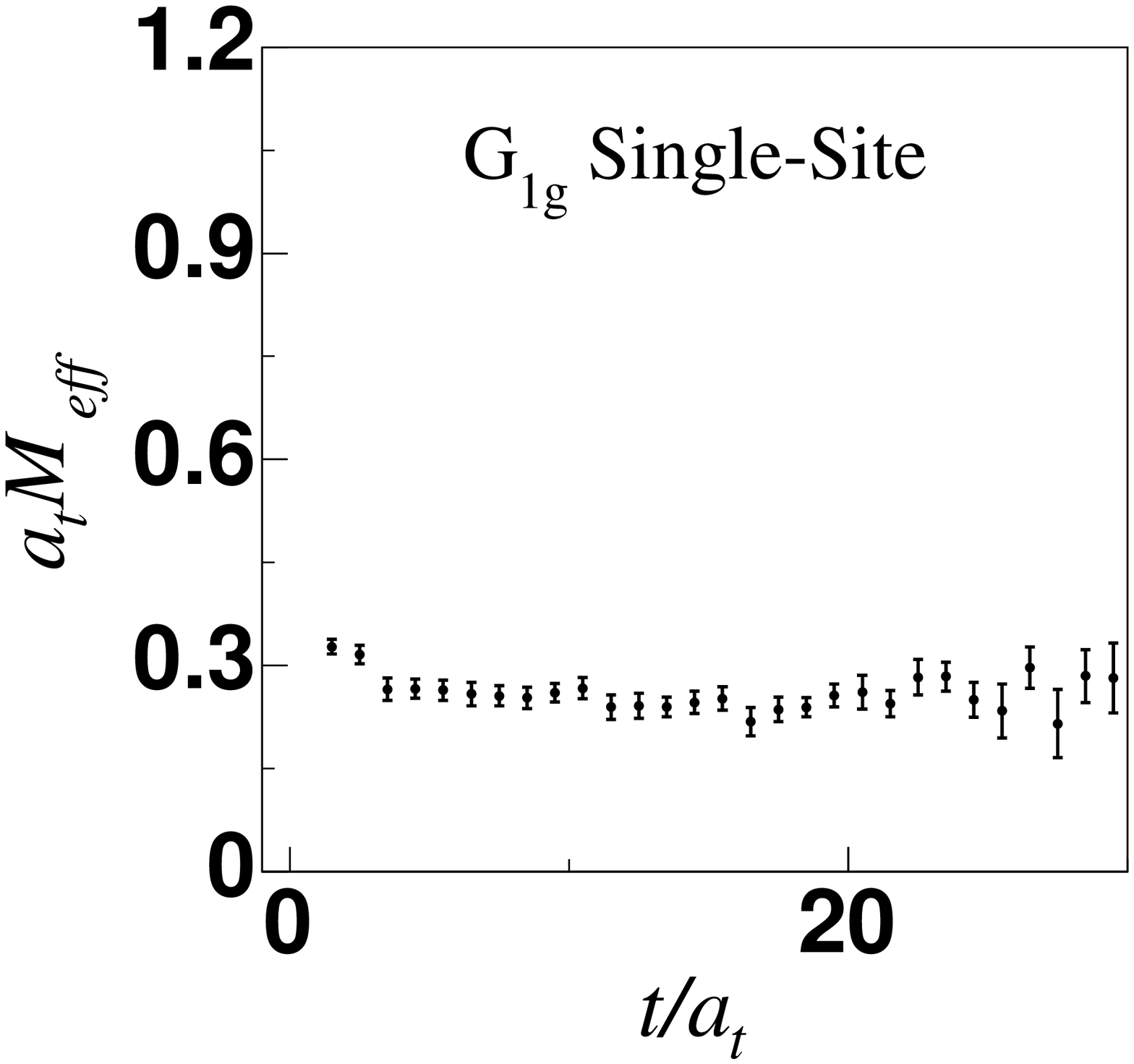,width=4cm,bb=0 50 567 520}
\psfig{file=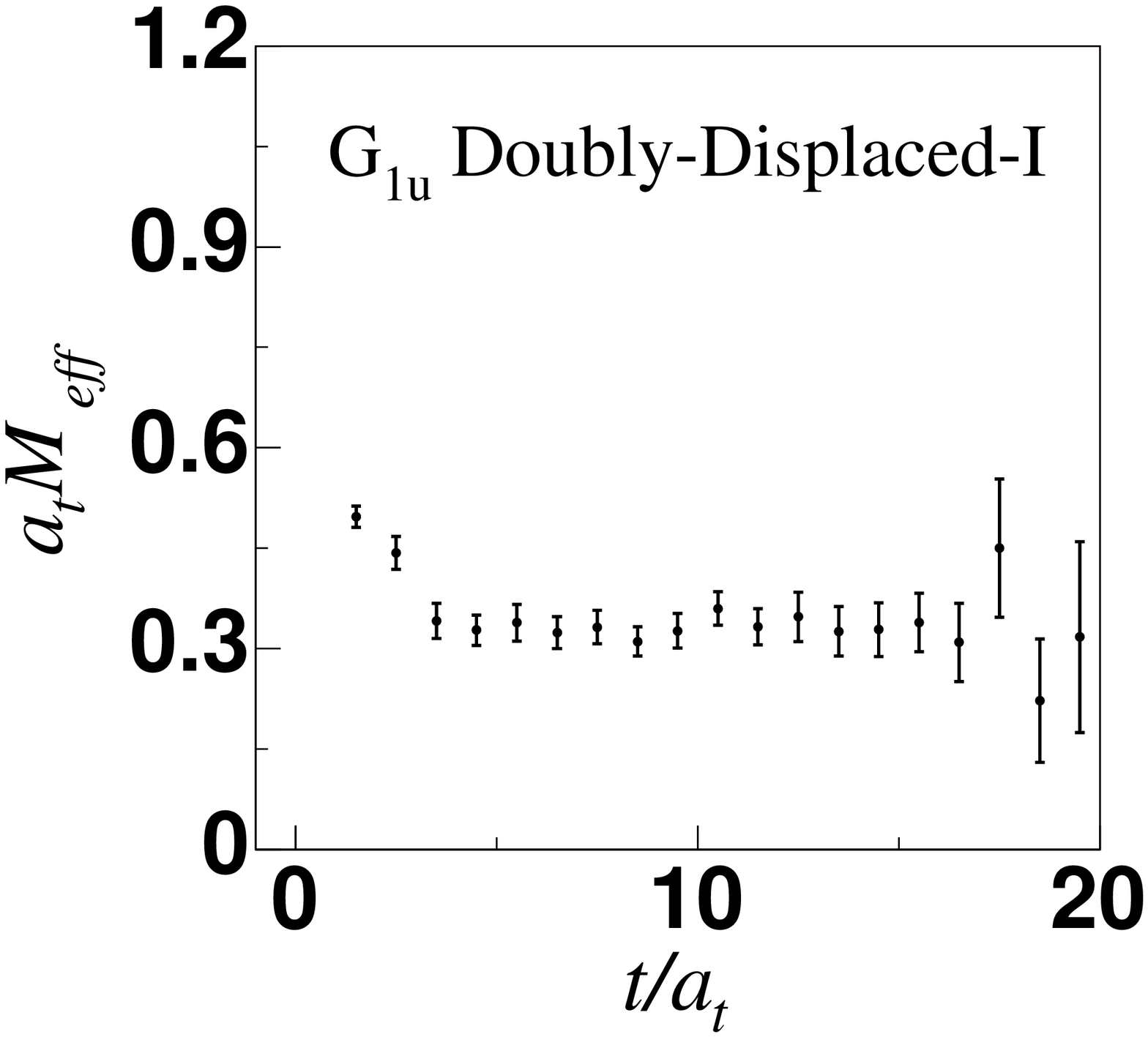,width=4cm,bb=0 50 567 520}
\psfig{file=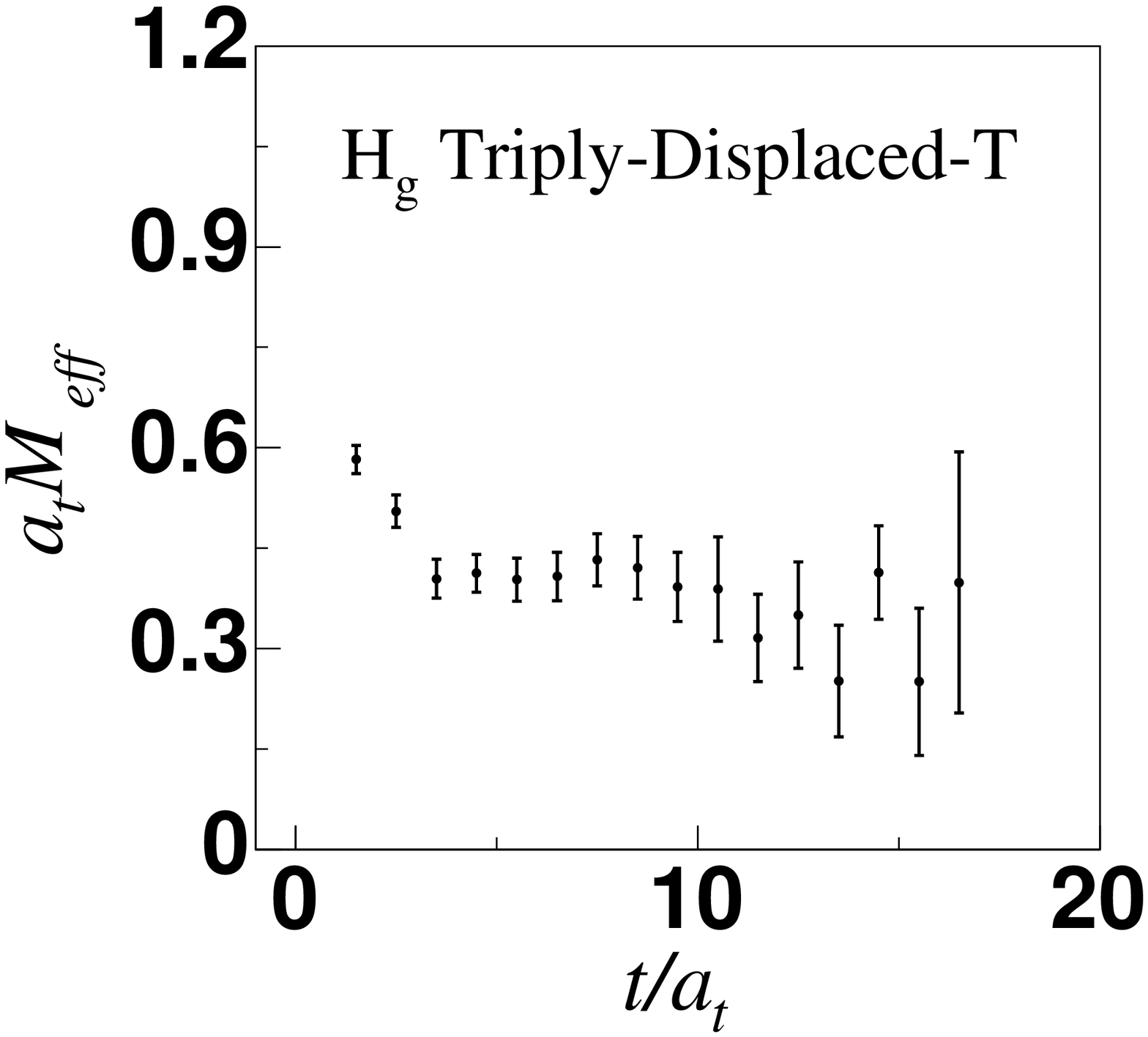,width=4cm,bb=0 50 567 520}}
\vspace*{8pt}
\caption{Effective mass plots for three particular nucleon operators:
a single-site operator in the $G_{1g}$ channel (left), a doubly-displaced-I
operator in the $G_{1u}$ channel (center), and a triply-displaced-T
operator in the $H_g$ channel. For a correlation function $C(t)$, the 
effective mass is defined by $m_{\rm eff}(t)=\ln(C(t)/C(t+1))$. These
results are based on 25 quenched configurations on a $12^3\times 48$
anisotropic lattice using the Wilson action and optimized quark-field
and link-variable smearing.
\label{fig:three}}
\end{figure}

Baryons are expected to be rather large objects, and hence, local operators
will not suffice.  Our approach to constructing extended operators is
to use covariant displacements of the quark fields.
Displacements in different directions are used to build up the appropriate
orbital structure, and displacements of different lengths can build up
the needed radial structure.  There are six different
spatial orientations that we use, shown in Fig~\ref{fig:two}.  The
singly-displaced operators are meant to mock up a diquark-quark coupling,
and the doubly-displaced and triply-displaced operators are chosen since
they favor the $\Delta$-flux and $Y$-flux configurations, respectively.

We have now finished optimizing the quark-field and gauge-field
smearing parameters, and have begun low-statistics runs to prune out
the ineffectual and overly-noisy operators.  Some sample effective
mass plots are shown in Fig.~\ref{fig:three}.  Our goal is to find
the smallest set of operators out of the several hundred we have
constructed which are useful for extracting some
number of lowest-lying states, and to push the technology
to maximize the number of excited-state energies which can be reliably
determined.  We shall report our findings in the near future.
This work was supported by the U.S.\ National Science Foundation
through grant PHY-0354982.


\begin{thebibliography}{00}    

\bibitem{cmichael}
C.~Michael, \textit{Nucl.\ Phys.\ B} \textbf{259}, 58 (1985).

\bibitem{luscherwolff}
 M.~L\"uscher and U.~Wolff, \textit{Nucl.\ Phys.\ B} \textbf{339}, 222 (1990).

\bibitem{dewitt}
 B.~DeWitt, \textit{Phys.\ Rev.\ } \textbf{103}, 1565 (1956).

\bibitem{wiese88}
 U.~Wiese, \textit{Nucl.\ Phys.\ B (Proc. Suppl.)} \textbf{9}, 609 (1989).

\bibitem{luscher91B}
 M.~L\"uscher, \textit{Nucl.\ Phys.\ B} \textbf{364}, 237 (1991).

\bibitem{rum95}
 K.~Rummukainen and S.~Gottlieb, \textit{Nucl.\ Phys.\ B} \textbf{450}, 397 (1995).

\bibitem{glueballs}
 C.~Morningstar and M.~Peardon, \textit{Phys.\ Rev.\ D} \textbf{60}, 034509 (1999).

\bibitem{jkm}
 K.~J.~Juge, J.~Kuti, and C.~Morningstar,
 \textit{Phys.\ Rev.\ Lett.\ } \textbf{90}, 161601 (2003). 

\bibitem{jkm2}
 K.J.~Juge, J.~Kuti, and C.J.~Morningstar,
 \textit{Phys.\ Rev.\ Lett.\ } \textbf{\bf 82}, 4400 (1999).

\bibitem{baryons}
 S.~Basak, R.~Edwards, G.T.~Fleming, U.M.~Heller, C.~Morningstar, D.~Richards,
 I.~Sato, and S.~Wallace, unpublished (hep-lat/0506029).
\end{thebibliography}
\end{document}